\documentclass{svjour2}                    

\smartqed  
\usepackage{graphicx}
\begin{document}

\title{Phase transition of meshwork models for spherical membranes
\thanks{This work was supported in part by a Grant-in-Aid for Scientific Research from Japan Society for the Promotion of Science.} 
}

\titlerunning{Phase transition of meshwork models}        

\author{Hiroshi Koibuchi}


\institute{Hiroshi Koibuchi \at
Department of Mechanical and Systems Engineering, Ibaraki National College of Technology, Nakane 866, Hitachinaka, Ibaraki 312-8508, Japan
 \\
              \email{koibuchi@mech.ibaraki-ct.ac.jp}           
}

\date{Received: date / Accepted: date}

\maketitle

\begin{abstract}
We have studied two types of meshwork models by using the canonical Monte Carlo simulation technique. The first meshwork model has elastic junctions, which are composed of vertices, bonds, and triangles, while the second model has rigid junctions, which are hexagonal (or pentagonal) rigid plates. Two-dimensional elasticity is assumed only at the elastic junctions in the first model, and no two-dimensional bending elasticity is assumed in the second model. Both of the meshworks are of spherical topology. We find that both models undergo a first-order collapsing transition between the smooth spherical phase and the collapsed phase. The Hausdorff dimension of the smooth phase is $H\!\simeq \!2$ in both models as expected. It is also found that $H\!\simeq \!2$ in the collapsed phase of the second model, and that $H$ is relatively larger than $2$ in the collapsed phase of the first model, but it remains in the physical bound, i.e., $H\!<\!3$.  Moreover, the first model undergoes a discontinuous surface fluctuation transition at the same transition point as that of the collapsing transition, while the second model undergoes a continuous transition of surface fluctuation. This indicates that the phase structure of the meshwork model is weakly dependent on the elasticity at the junctions.

\end{abstract}

\section{Introduction}\label{intro}
Two-dimensional curvature model of Helfrich and Polakov as well as that of Nambu and Goto has been extensively studied as a model of membranes and of strings from the viewpoints of two-dimensional differential geometry and statistical mechanics \cite{HELFRICH-1973,POLYAKOV-NPB1986,KLEINERT-PLB1986,Nambu-Select-1995}. Because of their two-dimensional nature, these models have a variety of surface shapes and the corresponding shape transformations or phase transitions, which can be considered to represent the complexity of real physical membranes \cite{NELSON-SMMS2004,Gompper-Schick-PTC-1994,Bowick-PREP2001}. 

Among the interesting topics on those models, the surface crumpling phenomenon is an old topic that has long been studied from several perspectives \cite{Peliti-Leibler-PRL1985,David-Guitter-EPL1988,PKN-PRL1988}. The phase transition of such crumpling phenomena is itself an interesting topic in biological and artificial membranes. Experimental investigations show that such phenomena can be seen in an artificial membrane \cite{CNE-PRL-2006}. In the string model context the path integration of the model describes the sum over surfaces in ${\bf R}^3$ \cite{Polyakov-contemp-1987,GSW-string-1987}, and therefore it seems that the summation technology changes depending on whether the surfaces are smooth or crumpled. The crumpling phenomena can also be seen in thin sheets. The universal structure was found in the formations of singularity in ridges and cones on those crumpled sheets \cite{CM-PRL1998,SCM-SCIE2000}. Moreover, if the transition is of second-order, the phenomena can be linked to a universal model for two-dimensional systems \cite{KacRaine-BLHWRIDLA-1987}.

Thus, extensive numerical studies on the phase transition have been made on triangulated surfaces \cite{KANTOR-NELSON-PRA1987,Baum-Ho-PRA1990,CATTERALL-NPBSUP1991,AMBJORN-NPB1993,KOIB-EPJB-2005}, including self-avoiding ones \cite{GREST-JPIF1991,BOWICK-TRAVESSET-EPJE2001,BCTT-PRL2001}.  
Recent numerical studies show that the conventional surface model has a first-order surface fluctuation transition on triangulated fixed-connectivity spheres \cite{KD-PRE2002,KOIB-PRE-2005,KOIB-NPB-2006}, and moreover, that the surface collapsing phenomena occur at the same transition point; the collapsing transition is considered to be a first-order transition. 

On the other hand, the conventional homogeneous surface model mentioned above can be extended to an inhomogeneous one by including a cytoskeletal structure \cite{BBD-BioPJ-1998,KOIB-JSTP-2007,KOIB-EPJB-2007}. In fact, the phase structure of skeleton models is partly understood \cite{KOIB-JSTP-2007,KOIB-EPJB-2007}. The numerical results show that the phase structure of the surface fluctuation phenomenon in the fixed-connectivity conventional surface model remains almost unchanged even when the compartmentalized structure was introduced \cite{KOIB-JSTP-2007}. On the contrary, in a dynamically triangulated fluid surface model the phase structure considerably changes if the free diffusion of vertices is confined inside the compartments, which are introduced as an inhomogeneous structure \cite{KOIB-EPJB-2007}. 

The compartmentalized models are those defined on a triangulated lattice with a cytoskeletal structure, which is a sublattice. Thus, the compartmentalized model is defined by using the lattice (= the surface) and the sublattice (= the cytoskeleton), because the one-dimensional bending energy is defined on the sublattice and the Gaussian bond potential is defined all over the lattice including the sublattice.

Therefore, it is natural to ask whether the surface shape is maintained only by the cytoskeletal structure. The problem which we consider is whether it is possible to eliminate the surface from the compartmentalized surface or not. If this is possible, then we are interested in whether the resulting model is well defined or not in the sense that the two-dimensional surface structure remains unchanged. Moreover, it is also interesting to see whether the phase structure remains unchanged when the surface is eliminated. 

Following to these considerations, we studied a meshwork model in \cite{KOIB-ICIC-2006} and reported some preliminary results on the phase structure. In this paper, we study two types of meshwork models including the one in \cite{KOIB-ICIC-2006} on relatively large sized surfaces. The first model in this paper is characterized by elastic junctions and is identical to the model in \cite{KOIB-ICIC-2006}, while the second model is characterized by rigid junctions, which are hexagonal (or pentagonal) rigid plates. 

We will see a first-order transition of surface-collapsing phenomena in both models. With respect to surface-fluctuations, the first model undergoes a first-order transition, while the second model a second-order transition. Thus, our results show that the phase structure of meshwork model is weakly dependent on the elasticity at the junctions. The Hausdorff dimension $H$ of the surface can be defined by using the mean square size $X^2$ and the total number of vertices $N$ such that $X^2\sim N^{2/H}$. It will also be shown that $H\!\simeq\!2$ in the smooth phase, and that $H$ remains in the physical bound, i.e., $H\!<\!3$, even in the collapsed phase in both models.

Thus, the first model in this paper has almost the same phase structure as that of the fixed-connectivity conventional surface model \cite{KOIB-PRE-2005}. On the contrary, the phase structure of the second model is slightly different from the fixed-connectivity conventional model because of the continuous transition of surface fluctuation.

\section{Models}\label{model}
The meshwork, as mentioned in the introduction, is constructed as compartments on the triangulated spherical surfaces by eliminating vertices, bonds, and triangles inside the compartments. Thus, the triangulated surfaces for constructing the meshwork are identical to the lattices for the surface models in \cite{KOIB-JSTP-2007,KOIB-EPJB-2007}. Therefore, the size of meshwork can be characterized by the expression similar to the one for those compartmentalized surface models in \cite{KOIB-JSTP-2007,KOIB-EPJB-2007}. 

Let $N$ be the total number of vertices, $N_S$ the total number of vertices on the chains, and $N_J$ the total number of junctions. Thus, the meshwork size can be denoted by $(N,N_S,N_J,L)$, where $L$ is the total number of bonds in a chain between the junctions; $L-1$ is the total number of vertices on the chain.

\begin{figure}[htb]
\unitlength 0.1in
\begin{picture}( 0,0)(  10,10)
\put(18,8.5){\makebox(0,0){(a) A meshwork with elastic }}%
\put(17.8,7.1){\makebox(0,0){junctions (model 1) }}%
\put(37,8.5){\makebox(0,0){(b) A meshwork with rigid }}%
\put(37.0,7.1){\makebox(0,0){junctions (model 2) }}%
\end{picture}%
\vspace{0.8cm}
\centering
\includegraphics[width=9.5cm]{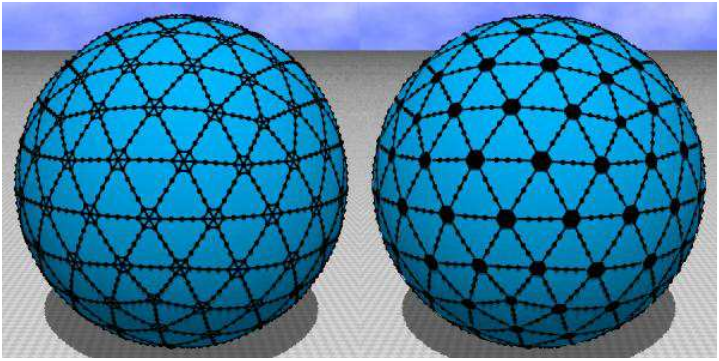}
\caption{(Color online) (a)A meshwork  with elastic junctions of size $(N,N_S,N_J,L)\!=\!(2562,1440,162,4)$, where $N$ is the total number of vertices, $N_S$ is the total number of vertices on the chains, $N_J$ is the total number of junctions, and $L$ is the length of chains between junctions, and  (b) a meshwork  with rigid junctions of size $(N,N_S,N_J,L)\!=\!(1602,1440,162,4)$, where $N\!=\!N_S\!+\!N_J$. The surfaces are drawn in order to visualize the meshwork clearly.} 
\label{fig-1}
\end{figure}
 Figure \ref{fig-1}(a) is a meshwork of size $(N,N_S,N_J,L)\!=\!(2562,1440,162,4)$ of the first model (denoted by model 1), where $N$ includes the vertices in the junctions, and $N_J$ includes $12$ pentagonal junctions. The small dots on the chains denote the vertices. The surface of the sphere is shown in the snapshot in order to clarify the meshwork.  Figure \ref{fig-1}(b) shows a meshwork of size $(N,N_S,N_J,L)\!=\!(1602,1440,162,4)$ for the second model (denoted by model 2), where $N$ is the total number of vertices, $N_S$, $N_J$, and $L$ are identical to those of model 1. The junctions are counted as vertices in $N$ and hence, $N$ is given also by $N\!=\!N_S\!+\!N_J$ in model 2. 

The meshwork is constructed as follows: Every edge of the icosahedron can be divided into $\ell$ pieces of uniform length, and then we have a triangulated surface of size $N_0\!=\!10\ell^2\!+\!2$ (= the total number of vertices on the surface). The compartmentalized structures are obtained by dividing $\ell$ further into $m$ pieces ($m\!=\!1,2,\cdots$). Thus, we have the chains of uniform length $L\!=\!(\ell /m)\!-\!2$ when $m$ divides $\ell$. The reason for the subtraction $-2$ is because of the junctions at the two end points of the chain. On the meshworks in Figs. \ref{fig-1}(a) and \ref{fig-1}(b), $\ell$ and $L$ are given by $\ell\!=\!24$ and $L\!=\!4$, and therefore $m\!=\!4$.

By using two integers $\ell$ and $m$, we have $N_J\!=\!10m^2+2$ and $N_S\!=\!30m(\ell\!-\!3m)$ in both models, and therefore $N\!=\!30m\ell\!-\!20m^2\!+\!2$ in model 1 and  $N\!=\!30m\ell\!-\!80m^2\!+\!2$ in model 2. Since the junctions are considered as vertices of the sublattice, we have the expression of $N_J\!=\!10m^2+2$. The total number of bonds in the sublattice is $3N_J\!-\!6$, and each bond contains $L\!-\!1$ vertices, then we have $N_S\!=\!(3N_J\!-\!6)(L-1)$. By using  $L\!=\!(\ell/m)\!-\!2$, we have the above expression of $N_S$. $N_J$ junctions in model 1 contain $7N_J\!-\!12$ vertices and therefore, we have $N\!=\!N_S\!+\!7N_J\!-\!12$ in model 1, whereas $N_J$ junctions in model 2 contain $N_J$ vertices and therefore, we have $N\!=\!N_S\!+\!N_J$ in model 2.

The total number of the compartments depends on the size $N$, and in fact, it increases with increasing $N$. However, the chain length $L$ can be chosen to be constant and independent of $N$. We fix the chain length $L$ to
\begin{equation}
\label{number-inside}
L=4,
\end{equation}
which corresponds to $n\!=\!10$, which is the total number of vertices inside a compartment of the surface model in Ref.\cite{KOIB-EPJB-2007}.

\begin{figure}[htb]
\unitlength 0.1in
\begin{picture}( 0,0)(  10,10)
\put(19,8.5){\makebox(0,0){(a) an elastic junction (model 1) }}%
\put(38,8.5){\makebox(0,0){(b) a rigid junction (model 2)  }}%
\end{picture}%
\vspace{0.5cm}
\centering
\includegraphics[width=4.5cm]{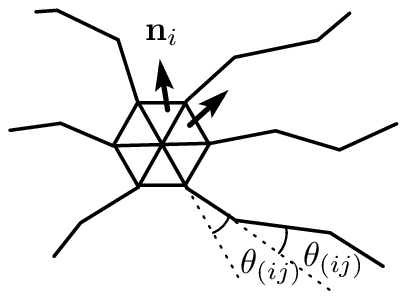}
\includegraphics[width=4.5cm]{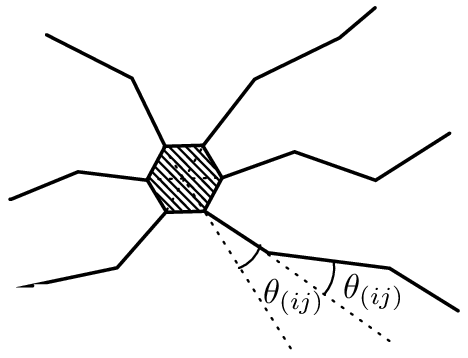}
\caption{(a) A hexagonal elastic junction of model 1, and (b) a hexagonal rigid junction of model 2. The unit normal vector ${\bf n}_i$ in (a) is defined on the triangle $i$ in the hexagon, and the angle $\theta_{(ij)}$ in $S_2$ is defined not only on the vertices of the chains but also on the corners of the junctions in (a) and (b). } 
\label{fig-2}
\end{figure}

Figure \ref{fig-2}(a) shows a hexagonal elastic junction of model 1. The unit normal vector ${\bf n}_i$ is defined on the triangle $i$ at the junctions of model 1, and the angle $\theta_{(ij)}$ is defined not only on the vertices of the chains but also on the corners of the junctions in both models. Figure \ref{fig-2}(b) shows a rigid junction of model 2.

The Hamiltonian $S(X)$ of model 1 is given by the linear combination of the Gaussian bond potential $S_1$, the one-dimensional bending energy $S_2$, and the two-dimensional bending energy $S_J$. $S_1$ is defined not only on the chains but also on the junctions, $S_2$ is defined only on the chains, and $S_J$ is defined only at the junctions:
\begin{eqnarray}
\label{Disc-Eneg-1} 
&& S(X)=S_1 + b S_2 + b_J S_J, \quad S_1=\sum_{(ij)} \left(X_i-X_j\right)^2,  \nonumber \\
&&S_2=\sum_{(ij)} \left[1-\cos \theta_{(ij)}\right], \quad S_J=\sum_{\langle ij \rangle} (1-{\bf n}_i \cdot {\bf n}_j), \qquad ({\rm model \; 1}),
\end{eqnarray} 
where $X_i (\in {\bf R}^3)$ denotes the three-dimensional position of the vertex $i$. $\sum_{(ij)}$ in $S_1$ is the sum over the bond $(ij)$ connecting the vertices $i$ and $j$ on the chains and on the junctions, and $\sum_{(ij)}$ in $S_2$ is the sum over bonds $i$ and $j$, which contain not only bonds in the chains but also bonds that connect the center and the corners of the junctions.  $\sum_{\langle ij \rangle}$ in $S_J$ is the sum over triangles $i$ and $j$, which share the center of the junction as the common vertex. The symbol $\theta_{(ij)}$ in $S_2$ is the angle between the bonds $i$ and $j$, and ${\bf n}_i$ in  $S_J$ is the unit normal vector of the triangle $i$ at the junctions as shown in Fig.\ref{fig-2}(a).

The coefficient $b$ is the one-dimensional bending rigidity (= coefficient of one-dimensional bending energy), which will be varied in order to see the phase structure, and $b_J$ is the two-dimensional bending rigidity at the junctions.  In this paper, $b_J$ is fixed to
\begin{equation}
\label{two-dim-rigidity}
b_J=5
\end{equation}
so that the junctions are sufficiently smooth. The value $b_J\!=\!5$ is relatively larger than the first-order transition point $b_c\!\simeq\! 0.77$ in the fixed-connectivity surface model \cite{KOIB-PRE-2005,KOIB-NPB-2006}. Therefore, the hexagonal or pentagonal junctions are almost flat even when the meshwork is in the crumpled phase at sufficiently small $b$.

Model 2 is defined on the meshwork with rigid junctions, such as the one shown in Fig.\ref{fig-1}(b).The Hamiltonian is given by the linear combination of the Gaussian bond potential $S_1$ and the one-dimensional bending energy $S_2$ such that
\begin{eqnarray}
\label{Disc-Eneg-2} 
&&S(X)=S_1 + b S_2, \quad S_1=\sum_{(ij)} \left(X_i-X_j\right)^2,\nonumber \\
&&S_2=\sum_{(ij)} \left[1-\cos \theta_{(ij)}\right],\qquad({\rm model \; 2}). 
\end{eqnarray} 
where $\sum_{(ij)}$ in $S_1$ is the sum over the bond $(ij)$ connecting the vertices $i$ and $j$, and $\sum_{(ij)}$ in $S_2$ is the sum over bonds $i$ and $j$, which contain not only bonds in the chains but also {\it virtual} bonds that connect the center and the corners of the rigid junctions; $S_2$ is defined on the vertices including the corners of the rigid junctions.

The size of the rigid junctions can be characterized by the edge length $R$, which is fixed to 
\begin{equation}
\label{Edge-Length} 
R=0.1\quad({\rm edge\; length\; of\; the \; rigid \; junctions} )
\end{equation}
in the simulations. The edge length $R\!=\!0.1$ is smaller than the mean bond length ($\simeq\! 0.707$); the bond length $0.707$ corresponds to that in the equilibrium configuration of surfaces without the rigid junctions, where the relation $S_1/N\!=\!1.5$ is satisfied. The rigid junctions in Fig.\ref{fig-1}(b) were drawn to have a size $R$ that is comparable to the mean bond length; the size $R$ in Fig.\ref{fig-1}(b) was drawn many times larger than $R\!=\!0.1$. As we will see later, the relation $S_1/N\!=\!1.5$ is almost satisfied in model 2.

The partition function $Z$ of model 1 and model 2 is defined by
\begin{equation} 
\label{Part-Func}
 Z = \int^\prime \prod _{i=1}^{N} d X_i \exp\left[-S(X)\right],
\end{equation} 
where $S(X)$ is the Hamiltonian, which is given in Eq.(\ref{Disc-Eneg-1}) or in Eq.(\ref{Disc-Eneg-2}). $\int^\prime $ denotes that the center of the meshwork is fixed in the integration. In model 1, the dynamical variables are integrated over $3N$-dimensional multiple integrations $\int^\prime \prod _{i=1}^{N} d X_i$, while in model 2 they are integrated over $3N_S$-dimensional integrations $\int^\prime \prod _{i=1}^{N_S} d X_i$ for the vertices and $6N_J$-dimensional integrations $\int^\prime \prod _{i=1}^{N_J} d X_i$ for the rigid junctions;
\begin{equation} 
\label{integration}
 \int^\prime \prod _{i=1}^{N} d X_i=\left(\int^\prime \prod _{i=1}^{N_S} d X_i\right)\left(\int^\prime \prod _{i=1}^{N_J} d X_i\right), \quad ({\rm model \;2}),
\end{equation} 
where $N\!=\!N_S\!+\!N_J$.

We must emphasize that the definitions of the models in this paper are quite different from those of the conventional surface models including the compartmentalized surface models such as the one in \cite{KOIB-JSTP-2007}: The conventional surface models are defined on the triangulated surfaces, which always include triangles (or plaquettes) even if the surface shape is maintained only by the skeletons.  On the contrary, the meshwork in this paper has no plaquettes except at the junctions in model 2 and is composed of the linear chains and the junctions only.

\section{Monte Carlo technique}\label{MC-Techniques}
In model 1, the integration of the dynamical variable $X$ is simulated by the random $3D$ shift from $X$ to $X^\prime\!=\!X\!+\!\delta X$, where $\delta X$ is randomly chosen in a small sphere. The new position $X^\prime$ is accepted with the probability ${\rm Min}[1,\exp(\delta S)]$, where $\delta S\!=\!S({\rm new})-S({\rm old})$. The radius of the small sphere $\delta X$ is fixed to certain constant value at the beginning of the simulations so that the acceptance rate is equal to about $50\%$.

The vertices on the junctions share an energy which is different from that shared by the vertices on the chains in model 1. For this reason, we adopt an additional random shift for the vertices on the junctions. The first step is a simultaneous $3D$ random translation of vertices on a junction, and the second step is a simultaneous $3D$ random rotation of those vertices. Both of the shifts are done under about $50\%$ acceptance rate. Note that these additional shifts of $X$ are not always necessary to simulate the integrations in model 1.

The integration of $X$ in model 2 is performed by a random $3D$ shift of the vertices on the chains and random $3D$ shifts of the rigid junctions, which respectively corresponds to the integrations in Eq.(\ref{integration}). The $3D$ shifts of the rigid junctions are done by a random $3D$ translation and a random $3D$ rotation of the rigid plates. All of the shifts of $X$ are done respectively under about $50\%$ acceptance rate.

A random number sequence called Mersenne Twister \cite{Matsumoto-Nishimura-1998} is used in the simulations. 

\section{Results of simulation}\label{Results}
\begin{figure}[htb]
\centering
\vspace{1.2cm}
\unitlength 0.1in
\begin{picture}( 0,0)(  10,10)
\put(14,35.3){\makebox(0,0){(a) Collapsed}}%
\put(15,34){\makebox(0,0){$b\!=\!2.92$}}%
\put(15,32.7){\makebox(0,0){(model 1)}}%
\put(24,35.3){\makebox(0,0){(b) Smooth}}%
\put(25,34){\makebox(0,0){$b\!=\!2.93$}}%
\put(25,32.7){\makebox(0,0){(model 1)}}%
\put(36,35.3){\makebox(0,0){(c)  Collapsed}}%
\put(37,34){\makebox(0,0){$b\!=\!2.59$}}%
\put(37,32.7){\makebox(0,0){(model 2)}}%
\put(46,35.3){\makebox(0,0){(d) Smooth}}%
\put(47,34){\makebox(0,0){$b\!=\!2.61$}}%
\put(47,32.7){\makebox(0,0){(model 2)}}%
\put(14,8.8){\makebox(0,0){(e) The section }}%
\put(24,8.8){\makebox(0,0){(f) The section}}%
\put(36,8.8){\makebox(0,0){(g) The section }}%
\put(46,8.8){\makebox(0,0){(h) The section}}%
\end{picture}%
\vspace{0.5cm}
\includegraphics[width=11.0cm]{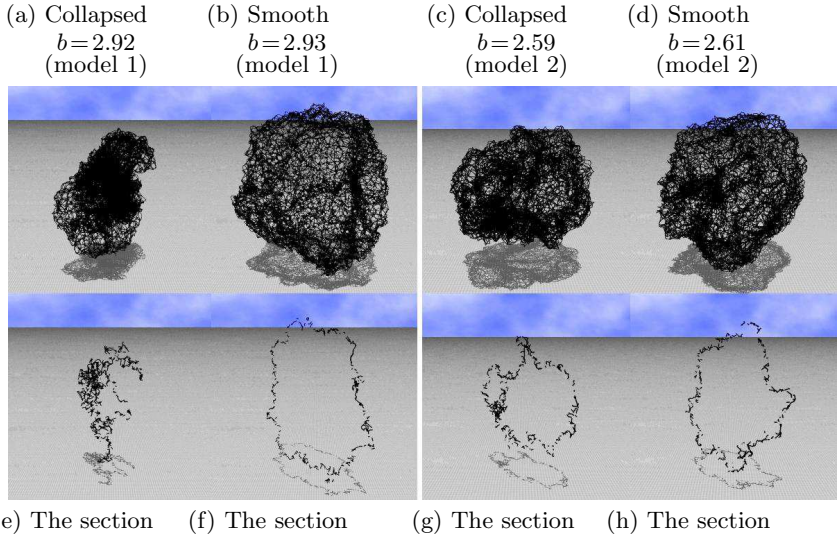}
\caption{(Color online) Snapshots of model 1 of size $(N,N_S,N_J,L)\!=\!(36002,20250,2252,4)$ obtained at (a) $b\!=\!2.92$ (collapsed phase), (b) $b\!=\!2.93$ (smooth phase), and snapshots of model 2 of size  $(N,N_S,N_J,L)\!=\!(22502,20250,2252,4)$ obtained at (c) $b\!=\!2.59$ (collapsed phase), (b) $b\!=\!2.61$ (smooth phase). The mean square size $X^2$, which is defined by Eq.(\ref{X2}), is (a) $X^2\!\simeq\!549$, (b) $X^2\!\simeq\!1563$,  (c) $X^2\!\simeq\!290$, and (d) $X^2\!\simeq\!832$. (e), (f), (g), and (h) are the meshwork sections.} 
\label{fig-3}
\end{figure}
First we show that both meshwork models have a two-dimensional surface structure at least in the smooth phase. Figures \ref{fig-3}(a) and \ref{fig-3}(b) are snapshots of model 1 surface of size $(N,N_S,N_J,L)\!=\!(36002,20250,2252,4)$ obtained at $b\!=\!2.92$ (collapsed phase) and $b\!=\!2.93$ (smooth phase), and  Figs.\ref{fig-3}(c) and \ref{fig-3}(d) are those of model 2 surface of size $(N,N_S,N_J,L)\!=\!(22502,20250,2252,4)$ obtained at $b\!=\!2.59$ (collapsed phase) and $b\!=\!2.61$ (smooth phase). The corresponding meshwork sections are shown in Figs.\ref{fig-3}(e)--\ref{fig-3}(h). The snapshots of model 1 were drawn in the same scale, which is slightly different from the scale for model 2. We see that the smooth state of model 1 in Fig.\ref{fig-3}(f) contains the empty space inside the meshwork just like the conventional surface model \cite{KOIB-PRE-2005}. However, the empty space is almost invisible in the crumpled state in Fig.\ref{fig-3}(e). On the contrary, both of the smooth state and the collapsed state of model 2 in Figs.\ref{fig-3}(g) and \ref{fig-3}(h) have empty spaces inside the meshworks.

\begin{figure}[htb]
\centering
\includegraphics[width=9.5cm]{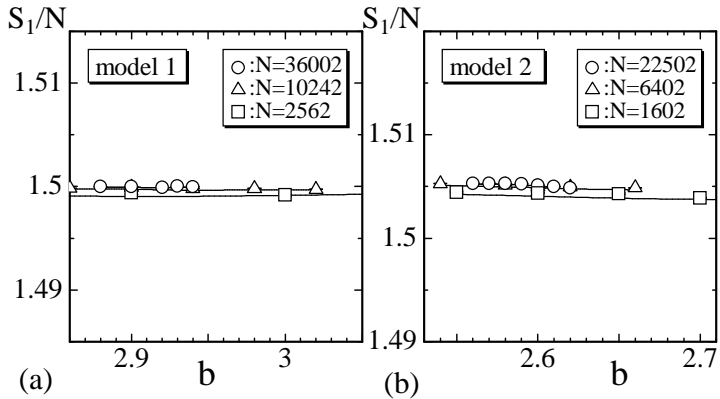}
\caption{The Gaussian bond potential $S_1/N$ versus $b$ of (a) model 1 and (b) model 2.  $S_1/N$ satisfies the predicted relation $S_1/N\!\simeq\!1.5$ in model 1, and it is slightly deviated from $1.5$ in model 2. The solid lines connecting the data were drawn by the multihistogram reweighting technique \cite{Janke-histogram-2002}. } 
\label{fig-4}
\end{figure}
The relation $S_1/N\!=\!3(N\!-\!1)/2N\!\simeq\!1.5$ is expected to be satisfied in model 1 because of the scale invariant property of the partition function \cite{Wheater-JPA-1994}. On the other hand, the expected relation can slightly be violated in model 2. This is because of the finite size of the rigid junctions, although the scale invariant property is still valid in model 2. Figures \ref{fig-4}(a) and \ref{fig-4}(b) show $S_1/N$ versus $b$ of model 1 and model 2. We can see from the figures that the expectations are fulfilled.

Therefore, we understand from the results shown in Fig.\ref{fig-4}(a) that the MC simulations for model 1 were successfully performed. We consider that the MC simulations for model 2 are as well, because the simulation technique for model 2 is almost identical to that for model 1. We find also from the results in Figs.\ref{fig-4}(a) and \ref{fig-4}(b) that the phase transition is not reflected in $S_1/N$ in contrast to the fluid surface model in Ref.\cite{KOIB-EPJB-2007}, where $S_1/N$ discontinuously changes at the transition point.

\begin{figure}[htb]
\centering
\includegraphics[width=9.5cm]{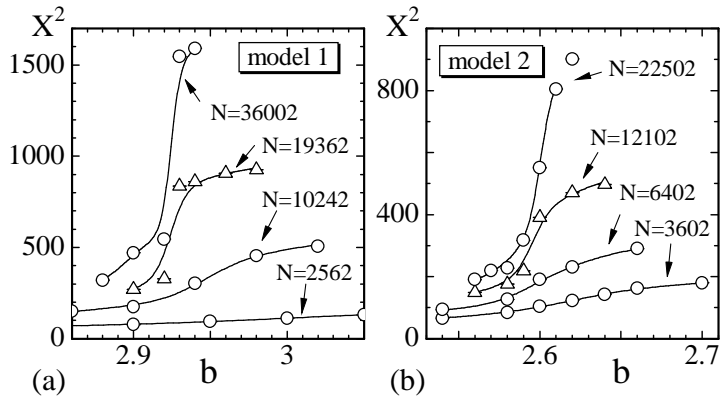}
\caption{The mean square size of $X^2$ versus $b$ of (a) model 1 and (b) model 2. The curves were drawn by the multihistogram reweighting technique.   } 
\label{fig-5}
\end{figure}
The mean square size $X^2$ is defined by 
\begin{equation}
\label{X2}
X^2={1\over N} \sum_i \left(X_i-\bar X\right)^2, \quad \bar X={1\over N} \sum_i X_i,
\end{equation}
where $\bar X$ is the center of the meshwork. $X^2$ represents the distribution of vertices in ${\bf R}^3$. Therefore, the meshwork is expected to be characterized by large $X^2$ (small $X^2$) at $b\!\to\! \infty(b\!\to\!0)$, where the meshwork is in a smooth (collapsed) state.

Figures \ref{fig-5}(a) and \ref{fig-5}(b) show $X^2$ versus $b$ of model 1 and model 2. The solid curves drawn on the data were obtained by the multihistogram reweighting technique \cite{Janke-histogram-2002}. We find from Fig.\ref{fig-5}(a) that $X^2$ changes almost discontinuously at an intermediate $b$. Although the multihistogram curves at the transition point appear to be smooth, the discontinuous change of $X^2$ is apparent in the $N\!=\!19362$ and $N\!=\!36002$ surfaces. This indicates the existence of a discontinuous phase transition of surface-collapsing phenomenon between the smooth spherical phase and the collapsed phase. On the contrary, $X^2$ of model 2 in Fig.\ref{fig-5}(b) appears to vary continuously against $b$, and therefore, we can not confirm the discontinuous nature of the transition in model 2 only from $X^2$ in Fig.\ref{fig-5}(b).  

\begin{figure}[htb]
\centering
\includegraphics[width=11cm]{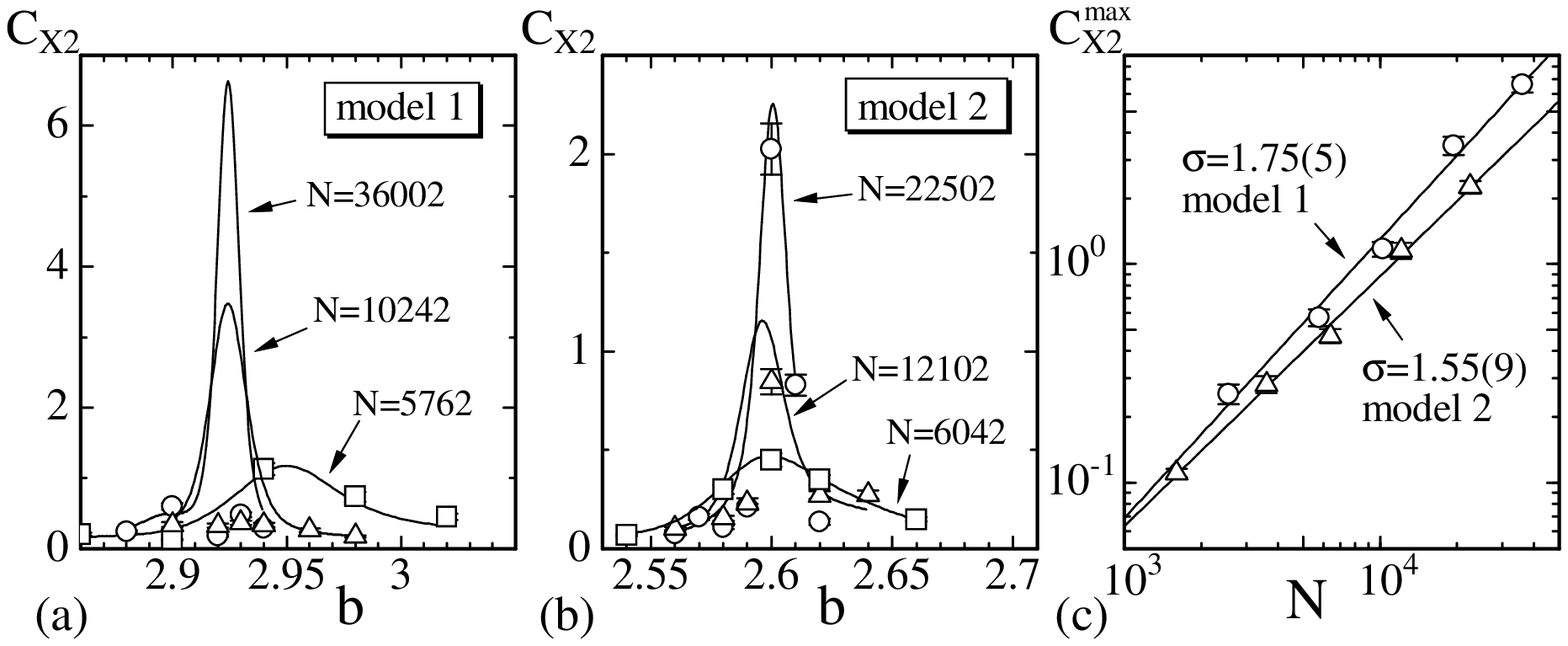}
\caption{The variance $C_{X^2}$ of $X^2$ versus $b$ of (a) model 1 and (b) model 2, and (c) log-log plots of the peak values $C_{X^2}^{\rm max}$ against $N$.} 
\label{fig-6}
\end{figure}
In order to see the order of the collapsing transition in both models, we compute the variance $C_{X^2}$ of $X^2$ by 
\begin{equation}
\label{specific-heat-X2}
C_{X^2} \!=\! {1\over N} \langle \; \left( X^2 \!-\! \langle X^2 \rangle\right)^2\rangle,
\end{equation}
which can reflect how large the skeleton size fluctuates. If the models undergo a collapsing transition, we can see an anomalous peak in $C_{X^2}$ at the transition point. We plot $C_{X^2}$ versus $b$ in Figs.\ref{fig-6}(a) and \ref{fig-6}(b), which were obtained in model 1 and model 2, respectively.

The solid curves in Figs.\ref{fig-6}(a) and \ref{fig-6}(b) were drawn by using the multihistogram reweighting technique as well as the corresponding $X^2$ curves in Figs.\ref{fig-5}(a) and \ref{fig-5}(b). We find in the curve of $C_{X^2}$ the expected anomalous peak, which apparently increases with increasing $N$. This indicates the existence of the collapsing transition in both models.

The order of the collapsing transition can be confirmed from the scaling of the peak values $C_{X^2}^{\rm max}$ such that 
\begin{equation}
\label{scaling-exponents-1}
C_{X^2}^{\rm max} \propto \left( N \right)^{\sigma_1},
\end{equation}
where $\sigma_1$ is a critical exponent. We show in Fig.\ref{fig-6}(c) the log-log plots of $C_{X^2}^{\rm max}$ against $N$, which were obtained from the curves in Figs.\ref{fig-6}(a) and \ref{fig-6}(b). The straight lines in Fig.\ref{fig-6}(c) were drawn by fitting the data to Eq.(\ref{scaling-exponents-1}). Thus, we have
\begin{equation}
\label{exponents-values-1}
\sigma_1=1.75\pm 0.05\; ({\rm model\; 1}),\quad \sigma_1=1.55\pm 0.09\;  ({\rm model\; 2}). 
\end{equation}
The finite-size scaling (FSS) theory indicates that the peak values $C_{X^2}^{\rm max}$ should scale according to $N^\sigma (\sigma\!=\!1)$ if the transition is of first order \cite{PRIVMAN-WS-1989,BINDER-RPP-1997,BNB-NPB-1993}. The exponents $\sigma_1$ in both models are larger than $1$, however, the observed scaling behavior in Fig.\ref{fig-6}(c) clearly reflects the anomalous behavior in $C_{X^2}$. Therefore,  the FSS analysis confirms that the model 2 undergoes a discontinuous collapsing transition, which was not always confirmed from the variation of $X^2$ against $b$ in Fig.\ref{fig-5}(b). Moreover, the FSS analysis of model 1 is considered to be consistent to the first-order collapsing transition indicated by the discontinuity of $X^2$ in Fig.\ref{fig-5}(a).

\begin{figure}[htb]
\centering
\includegraphics[width=9.5cm]{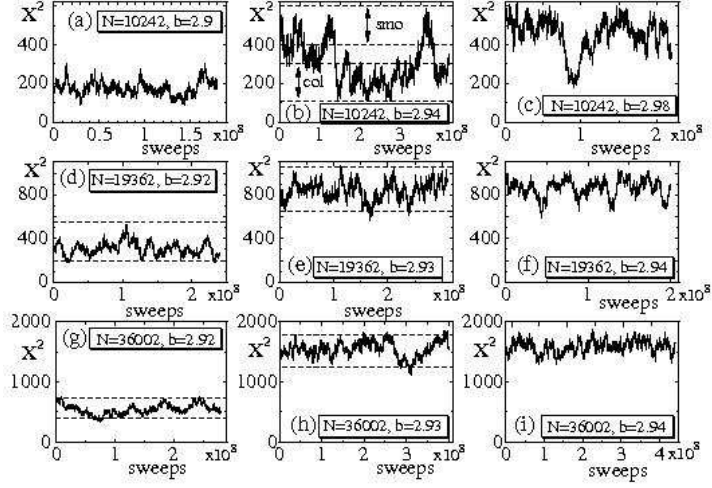}
\caption{The variation of $X^2$ versus MCS of model 1 surface of size $N\!=\!10242$ at (a) $b\!=\!2.9$, (b) $b\!=\!2.94$, and (c) $b\!=\!2.98$, and those of size $N\!=\!19362$ at (d) $b\!=\!2.92$, (e) $b\!=\!2.93$, and (f) $b\!=\!2.94$, and those of size $N\!=\!36002$ at (g) $b\!=\!2.92$, (h) $b\!=\!2.93$, and (i) $b\!=\!2.94$. Horizontal dashed lines denote $X^{2}_{\rm min}$ and  $X^{2}_{\rm max}$ for computing the mean value of $X^2$ in the collapsed phase and in the smooth phase, and $X^{2}_{\rm min}$ and  $X^{2}_{\rm max}$ are shown in Table \ref{table-1}. The symbols smo and col in (b) denote the smooth phase and the collapsed phase, respectively.} 
\label{fig-7}
\end{figure}
The discontinuous collapsing transition in both models can also be seen in the variations of $X^2$. Figures \ref{fig-7}(a)--\ref{fig-7}(i) show the variation of $X^2$ against MCS obtained at $b$ which are close to the transition point of model 1. The surfaces are of size $N\!=\!10242$,  $N\!=\!19362$, and  $N\!=\!36002$. We find that $X^2$ in Fig.\ref{fig-7}(a) remains in a lower value compared to that in Fig.\ref{fig-7}(c). The large fluctuation of $X^2$ in Fig.\ref{fig-7}(b) is consistent to the discontinuous transition between the collapsed phase and the smooth phase, which are respectively characterized by $X^2$ in Figs.\ref{fig-7}(a) and Fig.\ref{fig-7}(c). Contrary to $X^2$ of the $N\!=\!10242$ surface, we find no such fluctuation in $X^2$ on the $N\!=\!19362$ and $N\!=\!36002$ surfaces. After the surface is once trapped in the collapsed phase, it hardly changes to the smooth phase in those large sized surfaces. For this reason, a single variation of $X^2$ in the collapsed phase and two variations of $X^2$ in the smooth phase are shown in Figs.\ref{fig-7}(d)--\ref{fig-7}(f) and \ref{fig-7}(g)--\ref{fig-7}(i). Horizontal dashed lines denote $X^{2}_{\rm min}$ and  $X^{2}_{\rm max}$ for computing the mean value of $X^2$ in the collapsed phase and in the smooth phase. The Hausdorff dimension can be extracted from these mean values of $X^2$ and will be discussed below.  

\begin{figure}[htb]
\centering
\includegraphics[width=9.5cm]{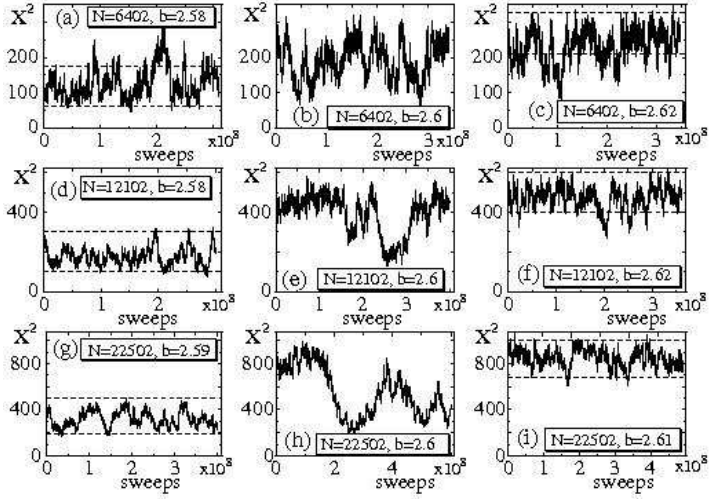}
\caption{The variation of $X^2$ versus MCS of model 2 surface of size $N\!=\!6402$ at (a) $b\!=\!2.58$, (b) $b\!=\!2.6$, and (c) $b\!=\!2.62$, and those of size $N\!=\!12102$ at (d) $b\!=\!2.58$, (e) $b\!=\!2.6$, and (f) $b\!=\!2.62$, and those of size $N\!=\!22502$ at (g) $b\!=\!2.59$, (h) $b\!=\!2.6$, and (i) $b\!=\!2.61$. Horizontal dashed lines denote $X^{2}_{\rm min}$ and  $X^{2}_{\rm max}$ for computing the mean value of $X^2$ in the collapsed phase and in the smooth phase. } 
\label{fig-8}
\end{figure}
The variations of $X^2$ of model 2 are shown in Figs.\ref{fig-8}(a)--\ref{fig-8}(i) obtained at $b$ which are close to the transition point of surfaces of size $N\!=\!6402$, $N\!=\!12102$, and $N\!=\!22502$. We find from the figures that the surfaces have two distinct states, which are respectively characterized by a large $X^2$ and a small $X^2$ just the same as those in Figs.\ref{fig-7}(a)--\ref{fig-7}(i) of model 1. However, jumps of $X^2$ can be seen in the variation in Figs.\ref{fig-8}(e) and \ref{fig-8}(h) even on such large surfaces. This indicates that the collapsing transition of model 2 is relatively weak compared to that of model 1. Horizontal dashed lines denote $X^{2}_{\rm min}$ and $X^{2}_{\rm max}$ for computing the mean value of $X^2$ in the collapsed phase and in the smooth phase, just as in Figs.\ref{fig-7}(a)-\ref{fig-7}(i).

The Hausdorff dimension $H$, which was already introduced in the final part of the Introduction, is obtained from the scaling relation 
\begin{equation}
\label{Hausdorff-Dim}
X^2 \propto N^{2/H}
\end{equation}
by using the mean values $X^2$ obtained in the collapsed phase and in the smooth phase. 

As indicated by the horizontal dashed lines in Figs.\ref{fig-7}(a)-\ref{fig-7}(i) and Figs.\ref{fig-8}(a)-\ref{fig-8}(i), a lower bound $X^{2 \;{\rm col}}_{\rm min}$ and an upper bound $X^{2 \;{\rm col}}_{\rm max}$ in the collapsed phase and those $X^{2 \;{\rm smo}}_{\rm min}$ and $X^{2 \;{\rm smo}}_{\rm max}$ in the smooth phase can be assumed for computing the mean values of $X^2$.  Table \ref{table-1} shows the assumed values for $X^{2}_{\rm min}$ and $X^{2}_{\rm max}$. Then, the mean values can be obtained from $X^2$ in the ranges $X^{2 \;{\rm col}}_{\rm min}\!<\!X^2\!<\!X^{2 \;{\rm col}}_{\rm max}$ and  $X^{2 \;{\rm smo}}_{\rm min}\!<\!X^2\!<\!X^{2 \;{\rm smo}}_{\rm max}$.  The symbols $b$(col) and $b$(smo) in Table \ref{table-1} denote that the bending rigidity where the sequence of $X^2$ was obtained for computing the mean value of $X^2$. The integers $(\ell,m)$ are those introduced in Section \ref{model} and characterize the size of the meshwork. 
\begin{table}[hbt]
\caption{ The lower bound $X^{2 \;{\rm col}}_{\rm min}$ and the upper bound $X^{2 \;{\rm col}}_{\rm max}$ for computing the mean value $X^2({\rm col})$ in the collapsed phase, and those $X^{2 \;{\rm smo}}_{\rm min}$ and $X^{2 \;{\rm smo}}_{\rm max}$ in the smooth phase. $b$(col) and $b$(smo) denote the bending rigidities where $X^2$ were obtained. }
\label{table-1}
\begin{center}
 \begin{tabular}{ccccccccc}
 \hline
model & $N$  & $(\ell,m)$ & $b$(col) & $X^{2 \;{\rm col}}_{\rm min}$ & $X^{2 \;{\rm col}}_{\rm max}$ & $b$(smo) &$X^{2 \;{\rm smo}}_{\rm min}$ & $X^{2 \;{\rm smo}}_{\rm max}$   \\
 \hline
  1 & 5762   & (36,6)  & 2.9  & 80  & 210  & 2.98 & 230  & 360  \\
  1 & 10242  & (48,8)  & 2.94 & 110 & 300  & 2.98 & 400  & 600  \\
  1 & 19362  & (66,11) & 2.92 & 200 & 550  & 2.93 & 650  & 1050  \\
  1 & 36002  & (90,15) & 2.92 & 400 & 730  & 2.93 & 1250 & 1800  \\
 \hline
  2  & 3602  & (36,6)  & 2.6  & 52  & 123  & 2.64 & 140 & 205  \\
  2  & 6402  & (48,8)  & 2.58 & 60  & 175  & 2.62 & 210 & 330  \\
  2  & 12102 & (66,11) & 2.58 & 100 & 300  & 2.62 & 400 & 600  \\
  2  & 22502 & (90,15) & 2.59 & 190 & 500  & 2.61 & 680 & 1000  \\
 \hline
\end{tabular} 
\end{center}
\end{table}

\begin{figure}[htb]
\centering
\includegraphics[width=9.5cm]{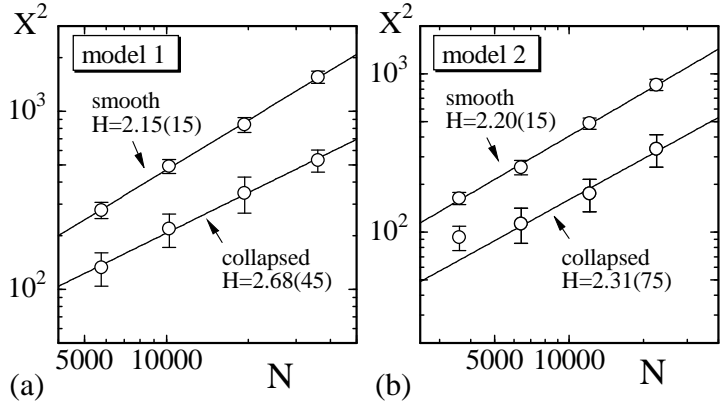}
\caption{ Log-log plots of $X^2({\rm col})$ and $X^2({\rm smo})$ against $N$ of (a) model 1 and (b) model 2. The straght lines were drawn by fitting the data to Eq.(\ref{Hausdorff-Dim}). The fitting in the collapsed phase in (b) was done by using the three largest data. } 
\label{fig-9}
\end{figure}

Figures \ref{fig-9}(a) and \ref{fig-9}(b) show the log-log plots of the mean values $X^2({\rm smo})$ and $X^2({\rm col})$ against $N$ of model 1 and model 2, respectively. The error bars are the standard deviations.

The straight lines on the figures were obtained by fitting the data to Eq.(\ref{Hausdorff-Dim}). The fittings were done by using four data in the figures except the case of the collapsed phase of model 2 in Fig.\ref{fig-9}(b). Thus, we have
\begin{eqnarray}
\label{H-values}
&&H^{\rm col}= 2.68\pm 0.45, \quad H^{\rm smo}= 2.15\pm 0.15 \quad ({\rm model\;1}), \nonumber \\
&&H^{\rm col}= 2.31\pm 0.75,\quad H^{\rm smo}= 2.20\pm 0.15 \quad ({\rm model\;2}).
\end{eqnarray}

We find from the results in Eq.(\ref{H-values}) that $H^{\rm smo}$ in the smooth phase of both models are almost identical to the topological dimension $H\!=\!2$, and that $H^{\rm col}$ of model 1 in the collapsed phase is different from $H^{\rm smo}$ in the smooth phase, while $H^{\rm col}$ and $H^{\rm smo}$ of model 2 are almost identical. $H^{\rm col}\!=\! 2.31(75)$ of model 2 implies that the meshwork is relatively smooth even in the collapsed phase, although the meshwork size discontinuously changes at the transition point. This is very similar to the case of the surface model with many holes in Ref.\cite{KOIB-PRE-2007-1}. To the contrary, $H^{\rm col}\!=\! 2.68(45)$ of model 1 implies that the meshwork is considerably collapsed in the collapsed phase. However, both $H^{\rm col}$ remain in the physical bound, i.e., $H^{\rm col}\!<\!3$. We should note that the results $H^{\rm col}$ of both models are in sharp contrast to that of the compartmentalized model in Ref.\cite{KOIB-JSTP-2007}, because the compartmentalized surface is completely collapsed in the collapsed phase. 

\begin{figure}[htb]
\centering
\includegraphics[width=9.5cm]{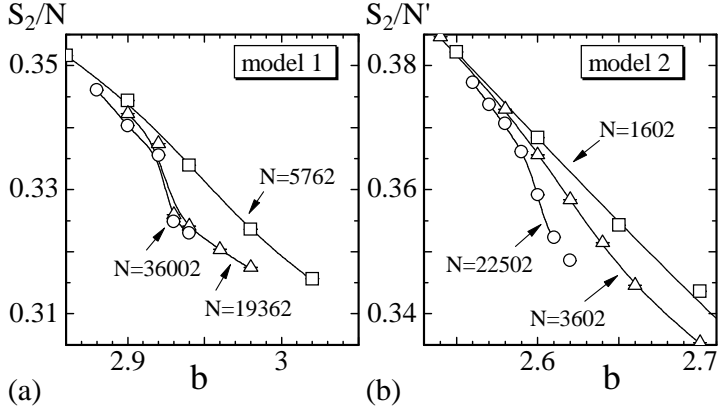}
\caption{ (a) The bending energy $S_2/N$ versus $b$ of model 1 and (b) the bending energy $S_2/N^\prime$ versus $b$ of model 2. $S_2$ is defined by Eq.(\ref{Disc-Eneg-1}) and  Eq.(\ref{Disc-Eneg-2}). $N$ and $N^\prime$ are the total number of vertices where $S_2$ is defined on the surfaces of model 1 and model 2, respectively.} 
\label{fig-10}
\end{figure}
We are interested also in the surface fluctuation phenomena; it is interesting to see whether or not the collapsing transition is accompanied by the surface fluctuation transition. The surface fluctuation phenomena can be characterized by the bending energy $S_2$ in Eq.(\ref{Disc-Eneg-1}) and Eq.(\ref{Disc-Eneg-2}), because $S_2$ is expected to be large (small) in the fluctuated (smooth) state in the meshwork model, just as the standard two-dimensional bending energy in the surface model. 

To see the surface fluctuations, we plot the bending energies $S_2/N$ and $S_2/N^\prime$ in Figs.\ref{fig-10}(a) and \ref{fig-10}(b), respectively. $N$ and $N^\prime$ are the total number of vertices where the bending energy is defined in model 1 and model 2, respectively. $N^\prime$ in model 2 is given by $N^\prime\!=\!N_S\!+\!6N_J\!-\!12\!=\!30m(\ell\!-\!m)$. Thus, we find that the variation of $S_2/N$ is discontinuous against $b$ in model 1; the jump can be seen in $S_2/N$ on the surfaces of $N\!=\!19362$ and $N\!=\!36002$, although the gap is relatively small compared to the value of $S_2/N$ itself. From the discontinuous change in $S_2/N$, we consider that the surface fluctuation transition is of first order in model 1. On the contrary, $S_2/N^\prime$ of model 2 in Fig.\ref{fig-10}(b) appears to vary continuously against $b$. 

\begin{figure}[htb]
\centering
\includegraphics[width=11cm]{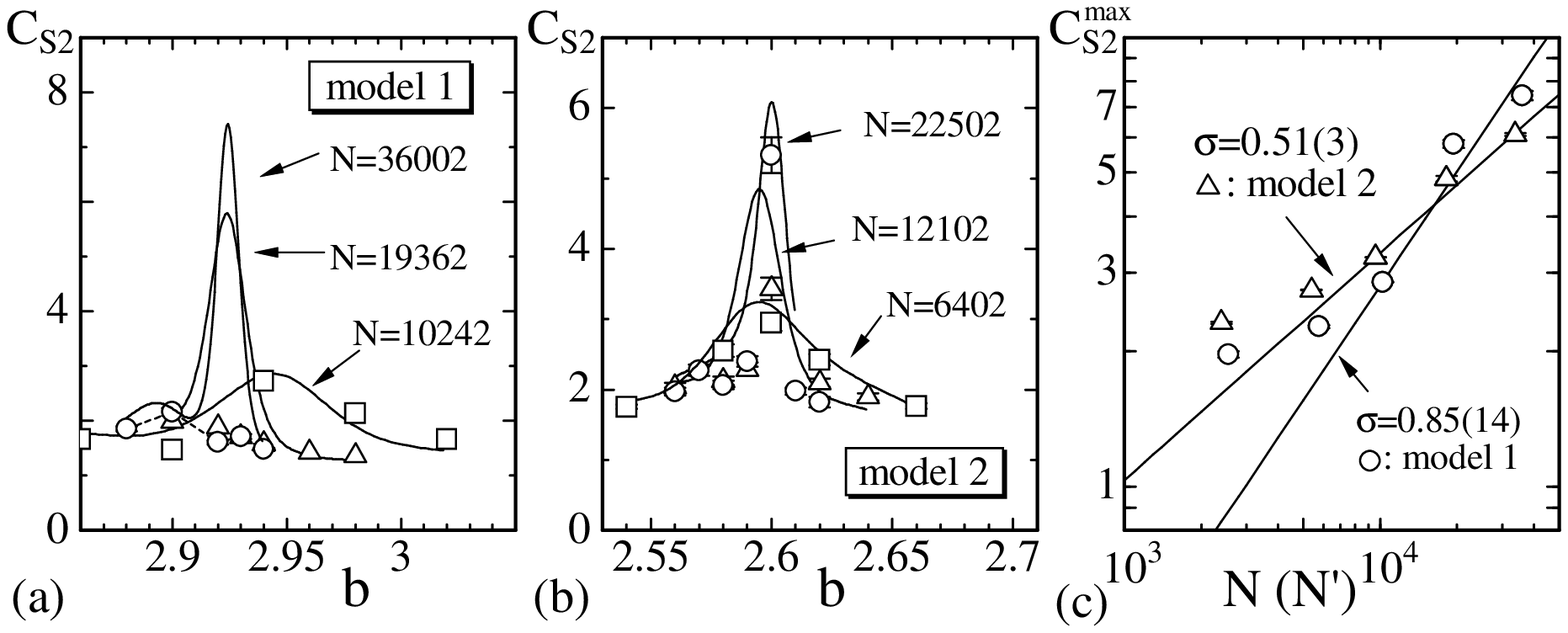}
\caption{The specific heat $C_{S_2}$ for $S_2$ versus $b$ of (a) model 1 and (b) model 2, and (c) log-log plots of the peak values $C_{S_2}^{\rm max}$ versus $N(N^\prime)$ of model 1 (model 2). The error bars on the symbol denote the statistical errors, which were obtained by the binning analysis. Solid curves in (a) and (b) were drawn by the multihistogram reweighting technique. The straight lines in (c) were drawn by fitting the largest three data (model 2) and the largest four data (model 1) to the expressions of Eq.(\ref{scaling-exponents-2}). } 
\label{fig-11}
\end{figure}
The specific heat for $S_2$ of model 1 is defined by
\begin{equation}
\label{specific-heat-S2}
C_{S_2} \!=\! {b^2\over N} \langle \; \left( S_2 \!-\! \langle S_2 \rangle\right)^2\rangle,
\end{equation}
which is the variance of $S_2$. $C_{S_2}$ of model 2 can be obtained by replacing $N$ by $N^\prime$ in Eq.(\ref{specific-heat-S2}). Just as $C_{X^2}$ that reflects the collapsing transition, $C_{S_2}$ can also reflect the surface fluctuation transition through its anomalous behavior. Figures \ref{fig-11}(a) and \ref{fig-11}(b) show $C_{S_2}$ versus $b$ of model 1 and model 2, respectively. The curves on the figures were obtained also by the multihistogram reweighting technique. We find an anomalous peak in $C_{S_2}$ of both models in Figs.\ref{fig-11}(a) and \ref{fig-11}(b). The peak values increase with increasing $N$ ($N^\prime$) and, therefore this indicates the existence of phase transition in both models.    

In order to see the order of the surface fluctuation transition, we show in Fig.\ref{fig-11}(c)  the log-log plots of the peak values $C_{S_2}^{\rm max}$ against $N(N^\prime)$; $N$ for model 1 and $N^\prime$ for model 2.  $C_{S_2}^{\rm max}$ were obtained from Figs.\ref{fig-11}(a) and \ref{fig-11}(b). The straight lines were drawn by fitting the data to 
\begin{equation}
\label{scaling-exponents-2}
C_{S_2}^{\rm max} \propto \left( N\right)^{\sigma_2}({\rm model \; 1}), \quad C_{S_2}^{\rm max} \propto \left( N^\prime\right)^{\sigma_2}({\rm model \; 2}),
\end{equation}
where $\sigma_2$ is a critical exponent. The fittings were done by using the largest four data in model 1 and the largest three data in model 2 in Fig.\ref{fig-11}(c). Thus, we have
\begin{equation}
\label{exponents-values-2}
\sigma_2=0.85\pm 0.14\; ({\rm model\; 1}), \quad \sigma_2=0.51\pm 0.03\; ({\rm model\; 2}). 
\end{equation}
The exponent $\sigma_2$ of model 1 can be seen as $\sigma_2\!\simeq\! 0.99$ within the error and, hence it is almost equal to $1$. Therefore, the scaling property of $C_{S_2}^{\rm max}$ seems consistent to the discontinuous transition indicated by the discontinuity of $S_2/N$ in Fig.\ref{fig-10}(a). On the contrary, the exponent $\sigma_2$ of model 2 is obviously smaller than $1$ and, therefore, this indicates that the surface fluctuation transition is considered to be of second order in model 2.

\begin{figure}[htb]
\centering
\includegraphics[width=9.5cm]{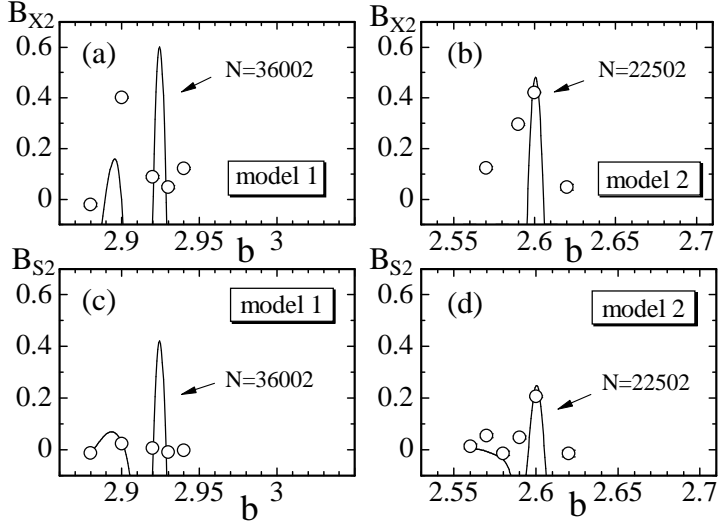}
\caption{The Binder quantity $B_{X^2}$ versus $b$ of (a) model 1 and (b) model 2, and $B_{S_2}$ versus $b$ of (c) model 1 and (d) model 2. 
} 
\label{fig-12}
\end{figure}
The order of the transitions can also be characterized by the Binder quantities $B_{X^2}$ and $B_{S_2}$ \cite{BINDER-ZFPB-1981}, which are defined by
\begin{equation}
\label{Binder-cumulants}
B_{X^2}=1-{\langle \left(X^2-\langle X^2\rangle\right)^4 \rangle \over 3\langle \left(X^2-\langle X^2\rangle\right)^2 \rangle^2 }, \quad
B_{S_2}=1-{\langle \left(S_2-\langle S_2\rangle\right)^4 \rangle \over 3\langle \left(S_2-\langle S_2\rangle\right)^2 \rangle^2 }.
\end{equation}
If the transition is of first order, then we should have $B\!=\!2/3$. In Figs.\ref{fig-12}(a)--\ref{fig-12}(d) we plot $B_{X^2}$ and $B_{S_2}$ of both models. The solid lines in the figures were drawn by the multihistogram reweighting technique.

We find in Figs.\ref{fig-12}(a) and \ref{fig-12}(b) that both $B_{X^2}$ and $B_{S_2}$ have a peak $B^{\rm max}$ at the transition point and that $B_{X^2}^{\rm max}$ is close to $2/3$ at the transition point of model 1, and also in Fig.\ref{fig-12}(b) that $B_{X^2}^{\rm max}$ is close to $2/3$  at the transition point of model 2. These results on the order of the collapsing transition in both models are consistent with the predictions by the discontinuity of $X^2$ in Fig.\ref{fig-5}(a) and by the finite size scaling analysis of $C_{X^2}^{\rm max}$ in Figs.\ref{fig-6}(c). It is also seen that $B_{S_2}^{\rm max}$ of model 1 in Fig.\ref{fig-12}(c) is relatively close to $2/3$ and that $B_{S_2}$ of model 2 in Fig.\ref{fig-12}(d) is relatively smaller than $2/3$, and therefore these results are also consistent with the results predicted by the observations in the FSS analyses of $C_{S_2}$.

\section{Summary and Conclusion}\label{Conclusion}
We have studied numerically two types of surface models defined on meshworks, which are constructed as a sublattice in a triangulated surface and are composed of linear chains and junctions. It was found that both models undergo a discontinuous collapsing transition between the smooth phase and the collapsed phase. The collapsed phases in both models are physical in the sense that $H\!<\!3$, where $H$ is the Hausdorff dimension. With respect to surface fluctuations, the first model undergoes a discontinuous transition, while the second model a continuous transition.

More precisely, two types of meshwork models of spherical topology were investigated by MC simulations for clarifying the phase structure, which is the order of the collapsing transition and the order of the surface fluctuation transition. Both models are defined on meshworks, which are composed of linear chains and junctions; no two-dimensional surface is included in the meshwork except at the junctions. The first model, denoted by model 1, is characterized by elastic junctions, which are composed of vertices, bonds and triangles, and the shape of the elastic junctions is of hexagonal and of pentagonal. The Hamiltonian of model 1 contains the Gaussian bond potential, the one-dimensional bending energy, and the two-dimensional bending energy at the elastic junctions. The second model, denoted by model 2, is characterized by rigid junctions, which are hexagonal rigid plates and pentagonal ones. The Hamiltonian of model 2 contains the Gaussian bond potential and the one-dimensional bending energy. 

The bending rigidity $b_J$ at the elastic junctions was fixed to $b_J\!=\!5$ in model 1, and the edge length $R$ of the rigid junctions was assumed to be $R\!=\!0.1$ in model 2. The compartment size was assumed to be $L\!=\!4$, which is the total number of bonds in a chain between the junctions. Thus, the compartment size can be negligible compared to the surface size if $N$ is sufficiently large. We used the lattices of size up to $N\!=\!36002$ in model 1 and those up to $N\!=\!22502$ in model 2.

We found that model 1 undergoes a first-order collapsing transition and a first-order surface fluctuation transition, and that model 2 undergoes a first-order collapsing transition and a second-order surface fluctuation transition. The smooth phase in both models is characterized by Hausdorff dimension $H\!\simeq \!2$, while the collapsed phases are slightly different from each other between the two models. The Hausdorff dimension $H^{\rm col}$ in the collapsed phase at the transition point of model 1 is relatively larger than the topological dimension of the surface, but it remains in the physical bound, i.e., $H\!<\!3$. To the contrary, $H^{\rm col}$ in the collapsed phase of model 2 is almost identical to $H^{\rm smo}$ in the smooth phase. This implies that the collapsed meshwork of model 2 has a two-dimensional surface structure just as in the smooth phase although the meshwork size discontinuously changes at the transition point.  

Our results in this paper also indicate that the phase structure of the meshwork model is dependent on the elasticity at the junctions. In fact, the surface fluctuation transition is of first-order in model 1 while that is of second-order in model 2.

Finally we comment on the relation between the models in this paper and the conventional fixed connectivity surface model \cite{KOIB-PRE-2005}. We consider that the meshwork models in this paper are almost identical to the conventional surface models because of the following three reasons: Firstly, both the conventional surface model and the meshwork models have a discontinuous collapsing transition. Secondly, both the conventional model and model 1 in this paper undergo a discontinuous transition of surface fluctuation. Thirdly, the collapsed phase of the meshwork models in this paper are physical, i.e., $H\!<\!3$, just as in the conventional model in \cite{KOIB-PRE-2005}, although no self-avoiding property is assumed in those models.

\vspace*{3mm}
\noindent
{\bf Acknowledgments}\\
This work was supported in part by a Grant-in-Aid for Scientific Research from Japan Society for the Promotion of Science.  


\end{document}